# Strongly coupled slow-light polaritons in one-dimensional disordered localized states

Jie Gao[1, 6*], Sylvain Combrie[2], Baolai Liang[3], Peter Schmitteckert[4], Gaelle Lehoucq[2], Stephane Xavier[2], Xinan Xu[1], Kurt Busch[5], Diana L. Huffaker[3], Alfredo De Rossi[2], and Chee Wei Wong[1*]

[1]*Columbia University, New York, NY 10027 USA*

[2]*Thales Research and Technology, Palaiseau, 91767, France*

[3]*University of California at Los Angeles, Los Angeles, CA 90095, USA*

[4]*Insititiue of Nanotechnology, Karlsruhe Institute of Technology, Eggenstein-Leopoldshafen, 76021, Germany*

[5]*Humboldt Universität zu Berlin, Institut für Physik, AG Theoretische Optik & Photonik, and Max Born Institute, 12489 Berlin,  Germany*

[6]*Department of Mechanical and Aerospace Engineering, Missouri University of Science and Technology, Rolla, MO 65409, USA*

**\*** gaojie@mst.edu; cww2104@columbia.edu

**Cavity quantum electrodynamics advances the coherent control of a single quantum emitter with a quantized radiation field mode, typically piecewise engineered for the highest finesse and confinement in the cavity field. This enables the possibility of strong coupling for chip-scale quantum processing, but till now is limited to few research groups that can achieve the precision and deterministic requirements for these polariton states. Here we observe for the first time coherent polariton states of strong coupled single quantum dot excitons in inherently disordered one-dimensional localized modes in slow-light photonic crystals. Large vacuum Rabi splittings up to 311 µeV are observed, one of the largest avoided crossings in the solid-state. Our tight-binding models with quantum impurities detail these strong localized polaritons, spanning different disorder strengths,**

**complementary to model-extracted pure dephasing and incoherent pumping rates. Such disorder-induced slow-light polaritons provide a platform towards coherent control, collective interactions, and quantum information processing.**

Solid-state cavity quantum electrodynamics (QED) serves as a critical resource for quantum information processing [1], guided by rapid demonstrations of sub-poissonian single photon sources [2,3], strong coupling in carefully engineered high quality factor ($Q$) cavities [4-9], precise positioning of single quantum dots with nanometer-scale accuracy [10], entanglement generation [11,12], spontaneous emission lifetime and coherent control [13-19]. The canonical strong coupling system involves a pre-designed optical cavity with finely tuned high $Q$ and small mode volume [20], providing an engineered platform to achieve strong coherent interactions between an individual exciton and a single photon. Alternatively, one-dimensional photonic crystal waveguides enable an increased local density of states at the slow-light band edge for enhanced light-matter interactions while affording ease of photon extraction, with recent Purcell-enhanced spontaneous emission demonstrations [21-24] and photon transport dynamics predictions [25]. With coherent interference and scattering at the slow-light band edge through inherent nanometer-scale fabrication fluctuations, high-$Q$ localized modes near the band edge were recently examined [26-30]. This enabled the remarkable observations of controlled spontaneous emission in single quantum dots through engineered disordered localized waveguide modes [21], though only in the weak coupling regime.

Here we demonstrate for the first time the strong exciton-photon coupling of single InAs excitons in inherently disordered waveguide localized states. Polaritons with large vacuum Rabi splittings up to 311 µeV are observed at the slow-light band edge. Ab initio tight-binding computations of the one-dimensional disordered waveguide verify both the existence of tightly-bounded narrow-linewidth localized modes near the band edge and coherent strongly coupled Rabi splitting of canonical two-level systems with optical resonances in the presence of a continuum of optical states. Quantum dissipative modeling retrieves the parameters for vacuum Rabi splitting, pure exciton dephasing, and incoherent cavity/exciton pumping rates. The disorder-induced strongly coupled system serves as a platform for coherent light-matter interactions and chip-scale quantum information processing.

**Results**

Figure 1(a) shows an example photonic crystal waveguide defined on a 120 nm thick GaAs membrane, with lattice period $a$ of 256 nm, hole radius $r$ of $0.244a$, and waveguide width $w$ of $0.98\sqrt{3}a$. InAs quantum dots in the membrane mid-plane are grown by molecular beam epitaxy with radiative spontaneous emission centered between 900 nm to 1000 nm (see Methods on the single quantum dot growth and device nanofabrication). To confirm the existence of single quantum dot exciton lines, micro-photoluminescence and Hanbury-Brown and Twiss measurements are performed (see Methods). Narrow single quantum dot exciton lines with linewidths ~ 90 to 120 µeV are observed with low (~ 1 µW) pump powers before the objective lens. Figure 1(b) illustrates an example antibunching measurement under 80 MHz Ti$^{3+}$:Sapphire laser excitation at 9 K with coincidence suppression to $g^{(2)}(0)$ of 0.28, revealing the sub-Poissonian photon statistics and single photon signature of the single quantum dots in the disordered membrane.

The inset of Figure 1(a) shows a close-up of the nanofabricated photonic crystal. The sample has inherent fabrication imperfections such as surface and sidewall roughness, non-uniform hole sizes and shape, and chamfered hole edges. Statistical analysis (see Supplementary Information) under high resolution scanning electron microscopy shows a 3.26% air hole radius fluctuation, and a root mean square (RMS) deviation of the edge roughness from a perfect circle at ~ 3.5 nm. The RMS of the hole center deviations from a perfect lattice along two principal $x$- and $y$-directions are 7.7 nm and 2.7 nm respectively when sampled across 56 holes. A roughness correlation length of 16.8 nm was observed in this sample.

Figure 2(a) shows the ab initio calculated band structure of the photonic crystal computed from 3D plane-wave expansion [31] for an ideal lattice. Our measurements are centered at the slow-light band edge of the higher-order guided mode (second mode, solid red) in the photonic band gap (fundamental guided mode in dashed blue), with a slow-light group index $n_g$ of approximately 25 to 60 at the van Hove singularity mode onset edge (highlighted by grey dashed rectangular region). The corresponding electromagnetic field distribution of this higher-order slow-light mode is shown in the inset. To examine the disorder and its corresponding localization, starting from the well-known Dicke-Hamiltonian [32], we derive a tight-binding Hamiltonian that describes photon propagation in an effectively one-dimensional disordered waveguide including coupling to a quantum impurity as [25]:

$$H = -J\sum_{x=1}^{M-1}(a_x^+ a_{x+1} + a_{x+1}^+ a_x) + \sum_{x=1}^{M}\varepsilon_x a_x^+ a_x + \frac{\Omega}{2}\sigma_z + V(a_{x_0}^+ \sigma_- + a_{x_0}\sigma_+).$$ Figure 2(b) first shows the computed spectral function of the localized modes (initially without the quantum impurity; coupling element $V = 0$) for the disorder strength $W$ at 0.8, where $W$ is defined as the site energy $\{\varepsilon_x\}$ uniformly random distributed over [-$W$/2, $W$/2] and lattice constant $a = 1$ is used in the model (see Supplemental Information). In addition to the ideal dispersion curve (white solid line, without disorder), sharp spectral features are observed especially near the slow-light mode edge, arising from multiple scattering of photons from the lattice structural disorder.

To examine the slow-light disorder localization further, in Figure 3(a) we illustrate the magnified energy- and momentum-resolved spectral functions for the selected $W$ at 3.0 scenario when $k/2\pi = 0.5$ and $0.46$. It shows that our low energy mode primarily couples to states corresponding to the slow-light band edge of the ideal structure. The observed states are localized in real space with exponentially-decaying wavefunctions, well-separated in energy, and with the scattering cross-section scaling quadratically with the local density of states in the middle of the spectrum for small disorder. In Figure 3(b), we next illustrate the obtained probability distributions of the localization lengths $\xi$ with three example disorder strengths $W$ at 0.2, 0.8 and 3.0 respectively, where the localization lengths decreased with increasing disorder strengths (see Supplemental Information). We emphasize that short localization lengths of a few lattice periods are observable in our computations, where the $W$ at 3.0 scenario matches well with the experimentally estimated localization lengths (from the strong coupling measurements) and with the large disorder found in our samples. The spectral functions and coherently coupled polariton states with an included quantum impurity are detailed in the Supplementary Information, where the spectral functions have been obtained via exact diagonalization. We also note that a single band model is examined here with constant hopping elements, though the extension to several bands and momentum-dependent coupling is possible, and specific lattice models can be further obtained via Wannier function approaches [33]. Encouraged by the theoretical computations, photoluminescence spectra are collected at 40 K from the vertically emitted radiation for two orthogonal polarizations as shown in Figure 3(c), with above band gap (632.8 nm) continuous-wave pump at 20 µW. Several localized cavity modes (indicated by black arrows) are observed near the slow-light band edge due to coherent random multiple scattering from disorder and are highly $y$-polarized with near absence in $x$-polarized spectrum. The quantum dot exciton lines (indicated by red arrows) and background emissions are present for both polarizations. In this scenario we note the ratios of the waveguide bandwidth to the full-width half-maximum of the localized mode resonances range from 400 to 650 at the slow-light band edge and compares well with our computed models (for *W* at 3.0) at approximately 400.

To further study the interaction between quantum dot excitons and waveguide localized modes, Figure 4(a) shows the temperature-dependent photoluminescence spectra at high excitation powers ~ 100 µW. The exciton lines strongly redshift with increasing temperature due to the modified band-gap energy. The frequency shift of localized cavity mode is smaller and the cavity resonance is around 915.25 nm. With the relatively high excitation power, the spectra (the top panel is an example spectrum at 20 K) consist of single exciton and biexciton states, localized modes and background quantum dot emissions. Figure 4(b) extracts the resonance frequencies of the single exciton ($X_a$) and the localized mode (LM) as a function of temperature. Figure 4(c) shows peak intensities of all the lines versus excitation power at 20 K, in which the localized mode intensity increases linearly with pump power, while the exciton lines have state-filling saturation behavior at high excitation powers.

We next examine at low excitation powers for low photon and exciton occupation numbers to enter into the strong coupling regime when the exciton transition is near or on-resonance with the disorder localized cavity mode. Figure 5(a) shows the two polariton peaks [$|p_\pm\rangle = (|X_y^0, 0_c\rangle \pm |g, 1_c\rangle)/\sqrt{2}$] (red and black open circles) with the avoided crossing at ~ 7 µW excitation with the tuned cold-finger temperature finely examined from 24 K to 32 K. Figure 5(b) illustrates the experimental spectra of the coherently coupled system in the open circles for temperatures around 27 K. The exciton - localized mode offsets 40 pm at 26 K, and the exciton transition ($X_a$) is redshifted by 230 pm and tuned to the longer wavelength side of the disorder-induced localized mode resonance at 29.5 K. Both single exciton and localized cavity mode observed here have linewidths ~ 90 µeV, corresponding to ~ 21 GHz decay rates and quality factor $Q \sim 15{,}000$ respectively. At 27 K, a large vacuum Rabi splitting of 311 µeV is observed when the exciton is on resonance with the localized mode at 915.255 nm. Blue dash lines represent the uncoupled localized mode and exciton transition as guidelines.

To examine the avoided crossing further in the presence of above band-gap incoherent pumping and exciton dephasing, we solve the master equation with the Hamiltonian

$$H = \frac{1}{2}\hbar\omega_x \sigma_z + \hbar\omega_c a^+ a + \hbar g(a^+ \sigma_- + a\sigma_+) \quad [1]$$

where $\sigma_+$, $\sigma_-$ and $\sigma_z$ are the pseudospin operators for the two-level quantum impurity with radiative transition frequency $\omega_x$, $a^+$ and $a$ are the photon creation and annihilation operators with localized cavity mode frequency $\omega_c$, and $g$ is the Rabi coupling frequency given by $g = \frac{\mu}{\hbar}\sqrt{\frac{\hbar\omega}{2\varepsilon V_{eff}}}$ where $\mu$ is the exciton dipole moment and $V_{eff}$ is the effective localized mode volume. The incoherent loss and pump terms [34] are included in the master equation of Lindblad form: $d\rho/dt = -i/\hbar[H,\rho] + L(\rho)$, with

$$\begin{aligned}L(\rho) = &\frac{\Gamma_x}{2}(2\sigma_-\rho\sigma_+ - \sigma_+\sigma_-\rho - \rho\sigma_+\sigma_-) + \frac{P_x}{2}(2\sigma_+\rho\sigma_- - \sigma_-\sigma_+\rho - \rho\sigma_-\sigma_+) \\ &+ \frac{\Gamma_c}{2}(2a\rho a^+ - a^+ a\rho - \rho a^+ a) + \frac{P_c}{2}(2a^+\rho a - aa^+\rho - \rho aa^+) \\ &+ \frac{\gamma_{dep}}{2}(\sigma_z\rho\sigma_z - \rho)\end{aligned} \quad [2]$$

Here $\Gamma_c$ is the localized mode decay rate, $\Gamma_x$ is the exciton decay rate, $P_x$ is the incoherent pumping rate of the exciton, $P_c$ is the incoherent pumping rate of the localized mode [35], and $\gamma_{dep}$ is the pure dephasing rate of the exciton. Effective decay rates are defined as $\Gamma_x' = \Gamma_x + 2\gamma_{dep} + P_x$, $\Gamma_c' = \Gamma_c - P_c$, and strong coherent coupling is observable if $g > (\Gamma_c'/4, \Gamma_x'/4)$ is achieved. We assume the collected light is mostly from the localized mode at steady-state with spectral function given by $\sim \int_0^\infty e^{-(\Gamma_r - i\omega)\tau} \langle a^+(t)a(t+\tau)\rangle d\tau$ where our system spectral resolution $\Gamma_r \sim 30$ μeV. From the theoretical model, we globally fit the entire set of temperature dependent spectra and illustrate the calculated peaks of the polariton states ($|p_+\rangle$ and $|p_-\rangle$) in Figure 5(a) (red and black solid lines) and Figure 5(b) (solid curves). The best experimental match is obtained for coupling strength $g$ at 39 GHz, pure dephasing rate $\gamma_{dep} \sim 10$ to 20 GHz, incoherent localized mode pumping rates $P_c \sim 5$ to 15 GHz and incoherent exciton pumping rates $P_x \sim 15$ GHz. $\Gamma_x$ of $\sim$ sub-GHz is used for the InAs exciton with nanosecond radiative recombination lifetimes.

**Discussion**

Due to the μW level above-band continuous-wave excitation and inevitable quantum dot interactions with the solid state environment, effects from a background continuum with multiple exciton lines, acoustic phonon scattering and spectral diffusion exist in the photoluminescence spectra. The fitted dephasing rate and incoherent pumping rates show their dependences on the temperature and localized cavity mode - exciton detuning, which is evidence for decoherence and incoherent pumping mediated by electron-phonon coupling [36,37]. An extra peak in the middle is observed in the photoluminescence, attributed to the unwanted and uncoupled excitonic transitions under the excitation powers of our above-band pump laser and in-plane cavity decay interference [38]. With the current quantum dot densities ($\sim 10^2$ per μm$^2$) in our GaAs membrane and the inherent random nature of our localized modes, we do not apply quantum dot spatial positioning for our sample so that there might be unwanted excitonic transitions in addition to the coherent spectra of the strongly coupled system. Laser induced quantum dot blinking [39] leads to the case that the exciton is occasionally not coupled with the localized mode and might also introduce a third extra peak in the photoluminescence. An extra Lorentzian peak in addition to the normal double polariton peaks is included to consider the above possibilities, and the fitted resonances and widths suggest that both the exciton and cavity could contribute to the extra peak, which reflects the complexity of the solid-state environment.

The coupling strength $g$ observed here is larger than previous values obtained from rigorously designed photonic crystal cavities with single quantum dots and in fact similar to the strongest coherent interaction reported so far [40], indicating that the spatial overlap between quantum dot and the localized cavity mode is remarkable through the intrinsic self-selection process. Among the plethora of disorder-induced localized modes near the slow-light edge, the quantum dot exciton preferably couples to the specific localized mode with the best spectral and spatial overlap. Assuming a reasonable spatial match between the exciton and the localized cavity mode, from the exciton-photon coupling strength we extract an upper bound to the localized polariton modal volume at $\sim 0.8\,(\lambda/n)^3$. By estimating the spatial extent of the localized polariton in $x$ and $z$ directions, we estimate a localization length of $\sim 1$ µm (which also corresponds with a few lattice periods in the statistical distributions modeled in Figure 3(b) for the $W$ at 3.0). Compared to recent studies of waveguide localized modes with engineered disorder [41] and theoretical suggestions of strong coupling possibilities [42], the localized modes observed and modeled here have short localization lengths ($\sim 1$ µm) and long loss lengths ($\sim 1.2$ mm), which enable the large $Q$ factor and strongly coupled polariton states observed in our experiments for the first time. Our Rabi-splitting measurements demonstrate that fabrication imperfections in the slow-light modes are sufficient to realize high $Q$ localized modes, achieve the polariton states, and serve as a platform for strong light-matter interactions.

We have demonstrated for the first time strong coherent polaritons through single InAs quantum dots excitons and one-dimensional disorder-induced localized modes in slow-light photonic crystals. Localized modes with short localization lengths and long loss lengths possess ~ 15,000 $Q$ factors and wavelength-scale confinement, while providing large spatial overlap with the embedded single excitons. The disorder-induced localized modes enable dramatically-large exciton-photon coupling strengths, and therefore strong coherent interactions, in semiconductor cavity QED via the always-present inherent fabrication imperfections. The observed vacuum Rabi splitting is up to 311 µeV, one of the largest observed in the solid-state. The strong coupling spectra are highly dependent on the experimental parameters and do not always appear as a symmetric mode splitting spectra. By comparing with theoretical models, tight-binding ab initio computations support the feasibility of strong coupling in disorder localized modes with narrow linewidths and short localization lengths (~ a few lattice periods). From the quantum equation model, contributions from pure dephasing, incoherent pumping, quantum dot blinking and undesired exciton transitions from finite temperatures are elucidated. The demonstrated slow-light exciton polaritons in an inherently disorder solid-state system provide a platform for bringing chip-scale quantum information processing and waveguide extraction-efficient secure communication modules closer to realization.

**Methods**

InAs quantum dots are embedded in the mid-plane of the 120 nm GaAs membrane with 620 nm AlGaAs sacrificial layer on a (001)-oriented GaAs wafer. The quantum dots are grown with density ~ $10^2$ per µm$^2$ dot by a solid-source molecular beam epitaxy via Stranski-Krastanov mode, and randomly distributed across the sample with ensemble photoluminescence centered between 900 nm to 1000 nm. The GaAs photonic crystal structures are patterned with a 100 kV electron-beam writer (NanoBeam *n*B3), inductively coupled reactive ion-etched, followed by a 40 seconds HF AlGaAs sacrificial layer wet-etch to release the suspended GaAs membranes.

Micro-photoluminescence and Hanbury-Brown and Twiss measurements are performed in a liquid helium flow cryostat. An above band-gap pump by $Ti^{3+}$:Sapphire 800 nm femtosecond laser or HeNe continuous-wave (CW) laser is focused to ~ 1 µm spot size using a 100× microscope objective (0.75 numerical aperture). The emitted photons are collected by the same objective, dispersed through a 1-m Horiba spectrometer, detected with a cooled charge-coupled device camera for photoluminescence spectra or two single photon counting avalanche photodiodes (APDs) for correlation measurements.

**References**

[1] H. Mabuchi and A. C. Doherty, Cavity quantum electrodynamics: coherence in context, *Science* **298**, 1372 (2002).

[2] S. Kako, C. Santori, K. Hoshino, S. Götzinger, Y. Yamamoto, and Y. Arakawa, A gallium nitride single-photon source operating at 200 K, *Nature Mater.* **5**, 887 (2006).

[3] M. Hijlkema, B. Weber, J. P. Specht, S. C. Webster, A. Kuhn, and G. Rempe, A single-photon server with just one atom, *Nature Phys.* **3**, 252 (2007).

[4] M. Nomura, N. Kumagai, S. Iwamoto, Y. Ota, and Y. Arakawa, Laser oscillation in a strongly coupled single-quantum-dot–nanocavity system, *Nature Phys.* **6**, 279 (2010).

[5] J. Kasprzak, S. Reitzenstein, E. A. Muljarov, C. Kistner, C. Schneider, M. Strauss, S. Höfling, A. Forchel, and W. Langbein, Up on the Jaynes–Cummings ladder of a quantum-dot/microcavity system, *Nature Mater.* **9**, 304 (2010).

[6] D. J. Alton, N. P. Stern, T. Aoki, H. Lee, E. Ostby, K. J. Vahala and H. J. Kimble, Strong Interactions of Single Atoms and Photons near a Dielectric Boundary, *Nature Phys.* **7**, 159 (2011).

[7] K. Hennessy, A. Badolato, M. Winger, D. Gerace, M. Atatüre, S. Gulde, S. Fält, E. L. Hu, and A. Imamoğlu, Quantum nature of a strongly coupled single quantum dot-cavity system, *Nature* **445**, 896 (2007).

[8] D. Englund, A. Faraon, I. Fushman, N. Stoltz, P. Petroff, and J. Vučković, Controlling cavity reflectivity with a single quantum dot, *Nature* **450**, 857 (2007).

[9] T. Yoshie, A. Scherer, J. Hendrickson, G. Khitrova, H. M. Gibbs, G. Rupper, C. Ell, O. B. Shchekin, and D. G. Deppe, Vacuum Rabi splitting with a single quantum dot in a photonic crystal nanocavity, *Nature* **432**, 200 (2004).

[10] A. Badolato, K. Hennessy, M. Atatüre, J. Dreiser, E. Hu, P. M. Petroff, and A. Imamoğlu, Deterministic coupling of single quantum dots to single nanocavity modes, *Science* **308**, 1158 (2005).

[11] A. Dousse, J. Suffczyński, A. Beveratos, O. Krebs, A. Lemaître, I. Sagnes, J. Bloch, P. Voisin, and P. Senellart, Ultrabright source of entangled photon pairs, *Nature* **466**, 217 (2010).

[12] A. Mulle, W. Fang, J. Lawall, and G. S. Solomon, Creating polarization-entangled photon pairs from a semiconductor quantum dot using the optical stark effect, *Phys. Rev. Lett.* **103**, 217402 (2009).

[13] E. B. Flagg, A. Muller, J.W. Robertson, S. Founta, D. G. Deppe, M. Xiao, W. Ma, G. J. Salamo, and C. K. Shih, Resonantly driven coherent oscillations in a solid-state quantum emitter, *Nature Phys.* **5**, 203 (2009).

[14] A. C. Arsenault, T. J. Clark, G. von Freymann, L. Cademartiri, R. Sapienza, J. Bertolotti, E. Vekris, S. Wong, V. Kitaev, I. Manners, R. Z. Wang, S. John, D. Wiersma, and G. A. Ozin, From colour fingerprinting to the control of photoluminescence in elastic photonic crystals, *Nature Mater.* **5**, 179 (2006).

[15] A. Faraon, I. Fushman, D. Englund, N. Stoltz, P. Petroff, and J. Vučković, Coherent generation of non-classical light on a chip via photon-induced tunnelling and blackade, *Nature Phys.* **4**, 859 (2008).

[16] S. Strauf, N. G. Stoltz, M. T. Rakher, L. A. Coldren, P. M. Petroff, and Dirk Bouwmeester, High frequency single-photon source with polarization control, *Nature Photonics* **1**, 704 (2007).

[17] R. Bose, D. Sridharan, H. Kim, G. Solomon, and E. Waks, Low-photon-number optical switching with a single quantum dot coupled to a photonic crystal cavity, *Phys. Rev. Lett.* **108**, 227402 (2012).

[18] M. D. Leistikow, A. P. Mosk, E. Yeganegi, S. R. Huisman, A. Lagendijk, and W. L. Vos, Inhibited spontaneous emission of quantum dots observed in a 3D photonic band gap, *Phys. Rev. Lett.* **107**, 193903 (2011).

[19] S. Noda, M. Fujita, and T. Asano, Spontaneous-emission control by photonic crystals and nanocavities, *Nature Photonics* **1**, 449 (2007).

[20] B.-S. Song, S. Noda, T. Asano, and Y. Akahane, Ultra-high-$Q$ photonic double-heterostructure nanocavity, *Nature Mater.* **4**, 207 (2005).

[21] L. Sapienza, H. Thyrrestrup, S. Stobbe, P. D. Garcia, S. Smolka, and P. Lodahl, Cavity quantum electrodynamics with Anderson-localized modes, *Science* **327**, 1352 (2010).

[22] E. Viasnoff-Schwoob, C. Weisbuch, H. Benisty, S. Olivier, S. Varoutsis, I. Robert-Philip, R. Houdré, and C. J. M. Smith, Spontaneous emission enhancement of quantum dots in a photonic crystal wire, *Phys. Rev. Lett.* **95**, 183901(2005).

[23] A. Laucht, S. Pütz, T. Günthner, N. Hauke, R. Saive, S. Frédérick, M. Bichler, M.-C. Amann, A. W. Holleitner, M. Kaniber, and J. J. Finley, A waveguide-coupled on-chip single-photon source, *Phys. Rev. X* **2**, 011014 (2012).

[24] T. Lund-Hansen, S. Stobbe, B. Julsgaard, H. Thyrrestrup, T. Sünner, M. Kamp, A. Forchel, and P. Lodahl, Experimental realization of highly efficient broadband coupling of single quantum dots to a photonic crystal waveguide, *Phys. Rev. Lett.* **101**, 113903 (2008).

[25] P. Longo, P. Schmitteckert, and K. Busch, Few-photon transport in low-dimensional systems: interaction-induced radiation trapping, *Phys. Rev. Lett.* **104**, 023602 (2010).

[26] M. Patterson, S. Hughes, S. Combrié, N.-V.-Quynh Tran, A. De Rossi, R. Gabet, and Y. Jaouën, Disorder-induced coherent scattering in slow-light photonic crystal waveguides, *Phys. Rev. Lett.* **102**, 253903 (2009).

[27] S. Mookherjea, J. S. Park, S.-H. Yang, and P. R. Bandaru, Localization in silicon nanophotonic slow-light waveguides, *Nature Photonics* **2**, 90 (2008).

[28] J. Topolancik, B. Ilic, and F. Vollmer, Experimental observation of strong photon localization in disordered photonic crystal waveguides, *Phys. Rev. Lett.* **99**, 253901(2007).

[29] M. Spasenović, D. M. Beggs, P. Lalanne, T. F. Krauss, and L. Kuipers, Measuring the spatial extent of individual localized photonic states, *Phys. Rev. B* **86**, 155153 (2012).

[30] V. Savona, Electromagnetic modes of a disordered photonic crystal, *Phys. Rev. B* **83**, 085301 (2011).

[31] S. G. Johnson and J. D. Joannopoulos, Block-iterative frequency-domain methods for Maxwell's equations in a planewave basis, *Optics Express* **8**, 173 (2001).

[32] R. H. Dicke, Coherence in spontaneous radiation processes, *Phys. Rev.* **93**, 99 (1954).

[33] K. Busch, C. Blum, A.M. Graham, D. Hermann, M. Köhl, P. Mack, and C. Wolff, The photonic Wannier function approach to photonic crystal simulations: status and perspectives, *J. Mod. Opt.* **58**, 365 (2011).

[34] F. P. Laussy, E. del Valle, and C. Tejedor, Strong coupling of quantum dots in microcavities, *Phys. Rev. Lett.* **101**, 083601 (2008).

[35] A. Laucht, N. Hauke, J. M. Villas-Bôas, F. Hofbauer, G. Böhm, M. Kaniber, and J. J. Finley, Dephasing of exciton polaritons in photoexcited InGaAs quantum dots in GaAs nanocavities, *Phys. Rev. Lett.* **103**, 087405 (2009).

[36] M. Calic, P. Gallo, M. Felici, K. A. Atlasov, B. Dwir, A. Rudra, G. Biasiol, L. Sorba, G. Tarel, V. Savona, and E. Kapon, Phonon-mediated coupling of InGaAs/GaAs quantum-dot excitons to photonic crystal cavities, *Phys. Rev. Lett*. **106**, 227402 (2011).

[37] S. Hughes, P. Yao, F. Milde, A. Knorr, D. Dalacu, K. Mnaymneh, V. Sazonova, P. J. Poole, G. C. Aers, J. Lapointe, R. Cheriton, and R. L. Williams, Influence of electron-acoustic phonon scattering on off-resonant cavity feeding within a strongly coupled quantum-dot cavity system, *Phys. Rev. B* **83**, 165313 (2011).

[38] S. Hughes and P. Yao, Theory of quantum light emission from a strongly-coupled single quantum dot photonic-crystal cavity system, *Optics Express* **17**, 3322 (2009).

[39] A. Reinhard, T. Volz, M. Winger, A. Badolato, K. J. Hennessy, E. L. Hu, and A. Imamoğlu, Strongly correlated photons on a chip, *Nature Photonics* **6**, 93 (2012).

[40] M. Winger, A. Badolato, K. J. Hennessy, E. L. Hu, and A. Imamoğlu, Quantum dot spectroscopy using cavity quantum electrodynamics, *Phys. Rev. Lett.* **101**, 226808 (2008).

[41] S. Smolka, H. Thyrrestrup, L. Sapienza, T. B. Lehmann, K. R Rix, L. S. Froufe-Pérez, P. D García, and P. Lodahl, Probing the statistical properties of Anderson localozation with quantum emitters, *New J. Phys.* **13**, 063044 (2011).

[42] H. Thyrrestrup, S. Smolka, L. Sapienza, and P. Lodahl, Statistical theory of a quantum emitter strongly coupled to Anderson-localized modes, *Phys. Rev. Lett.* **108**, 113901 (2012).

**Acknowledgements**

We thank Stefan Strauf, Andrzej Veitia, and Dirk Englund for helpful discussions. Financial support was provided by NSF DMR (Materials World Network 1108176), NSF ECCS (CAREER 0747787), NSF DGE (IGERT 1069240), DOD NSSEFF (N00244-09-1-0091), Deutsche Forschungsgemeinschaft (DFG), and the State of Baden-Württemberg through the DFG-Center for Functional Nanostructures (CFN) within subproject B2.10.

**Author Contributions**

Experiments were designed and carried out by J.G. and C.W.W. Data were analyzed and interpreted by J.G., C.W.W, and X.X. The samples were grown by B.L. and D.L.H. and devices fabricated by S.C., G.L., S.X. and A.D. The tight binding model result was calculated by P.S. and K.B. All authors subsequently contributed to manuscript writing.

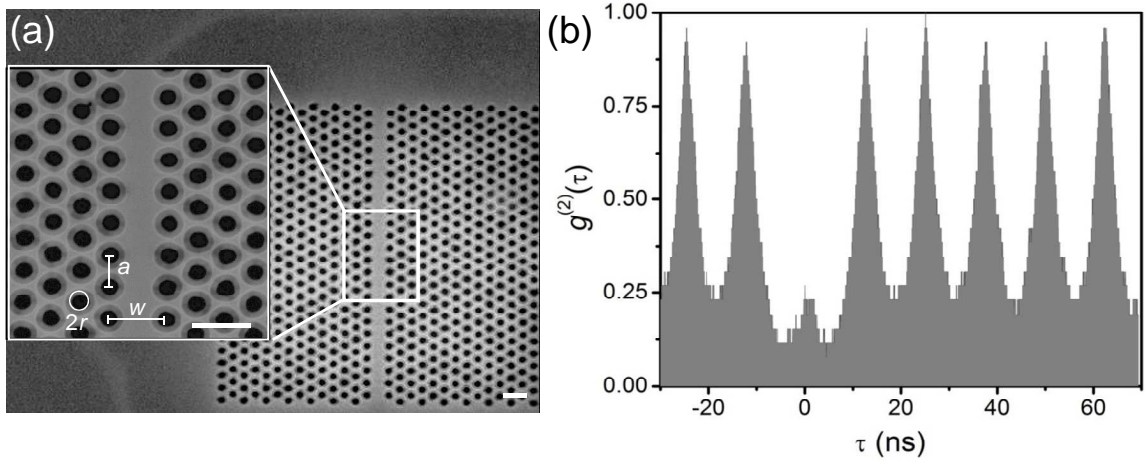

**Figure 1 | Disordered one-dimensional photonic crystal membrane with single photon emission and antibunching. a,** Scanning electron micrograph (SEM) of a nanofabricated disorder-inherent photonic crystal. Scale bars: 500 nm. **b,** Hanbury-Brown and Twiss antibunching of embedded single InAs quantum dot and single photon measurements under pulsed $Ti^{3+}$:Sa 800 nm laser excitation, with $g^{(2)}(0)$ at 0.28 in this example of waveguide sub-Poissonian single photon source.

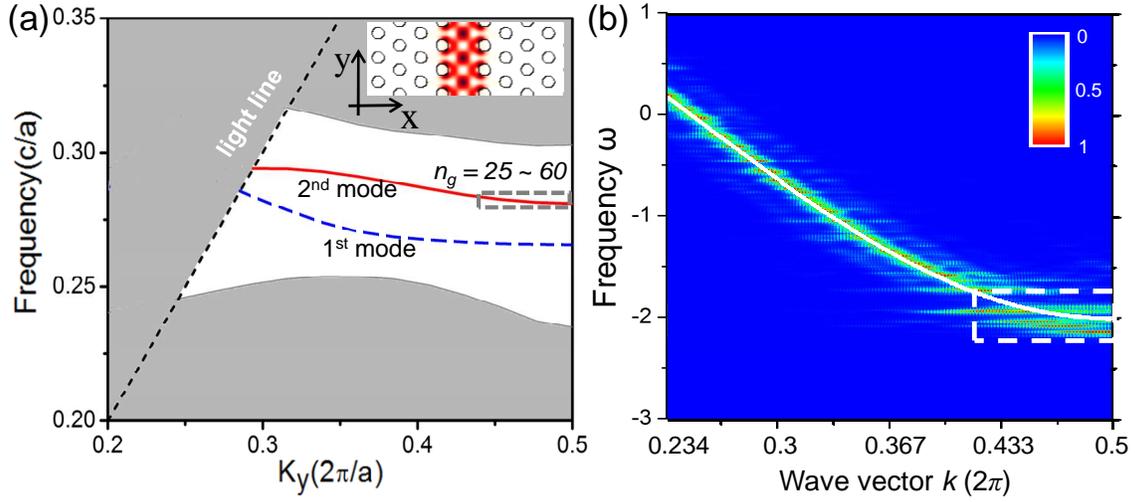

**Figure 2 | Band structure and spectral functions of the one-dimensional disorder localized modes. a,** Band structure for an ideal PhC waveguide with the designed parameters. Red and blue lines indicate the higher-order and fundamental waveguide modes. The grey dashed rectangular box indicates the slow-light band edge region (with group index $n_g$ 25 ~ 60) of higher-order waveguide mode, and the inset shows the electric field energy $|E|^2$ distribution at $K_y=\pi/a$. **b,** Spectral function of the disordered 1D waveguide described by the tight-binding chain model (band structure of an ideal structure is plotted in white solid line) with disorder strength $W = 0.8$. The white dashed rectangular box represents the slow-light band edge region.

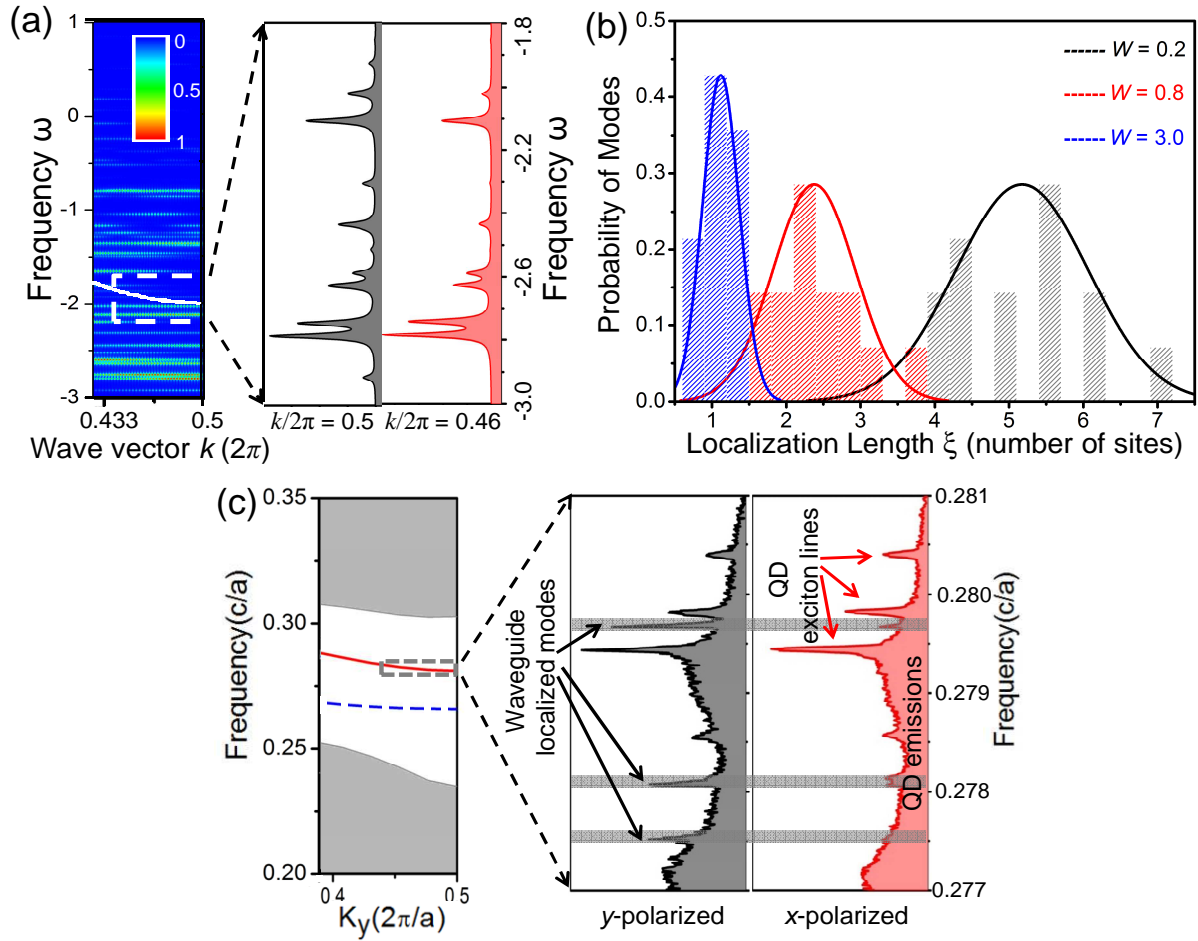

**Figure 3 | Cryogenic photoluminescence of the slow-light disorder localized modes and tight-binding computed resonant mode spectral functions. a,** Resonance modes in spectral functions of the disordered one-dimensional tight-binding waveguide at $k/2\pi = 0.5$ and $k/2\pi = 0.46$ when disorder strength $W = 3.0$. **b,** The probability distributions of the localization lengths for three disordered one-dimensional waveguides with $W$ of 0.2 (black), 0.8 (red) and 3.0 (blue). **c,** $y$-polarized and $x$-polarized photoluminescence spectra of localized modes from the fabricated disordered photonic crystal near slow-light region (peaks indicated by black arrows), quantum dot exciton lines (peaks indicated by red arrows) and background quantum dot emissions around the slow-light region when T = 40 K with high excitation power ~ 20 μW.

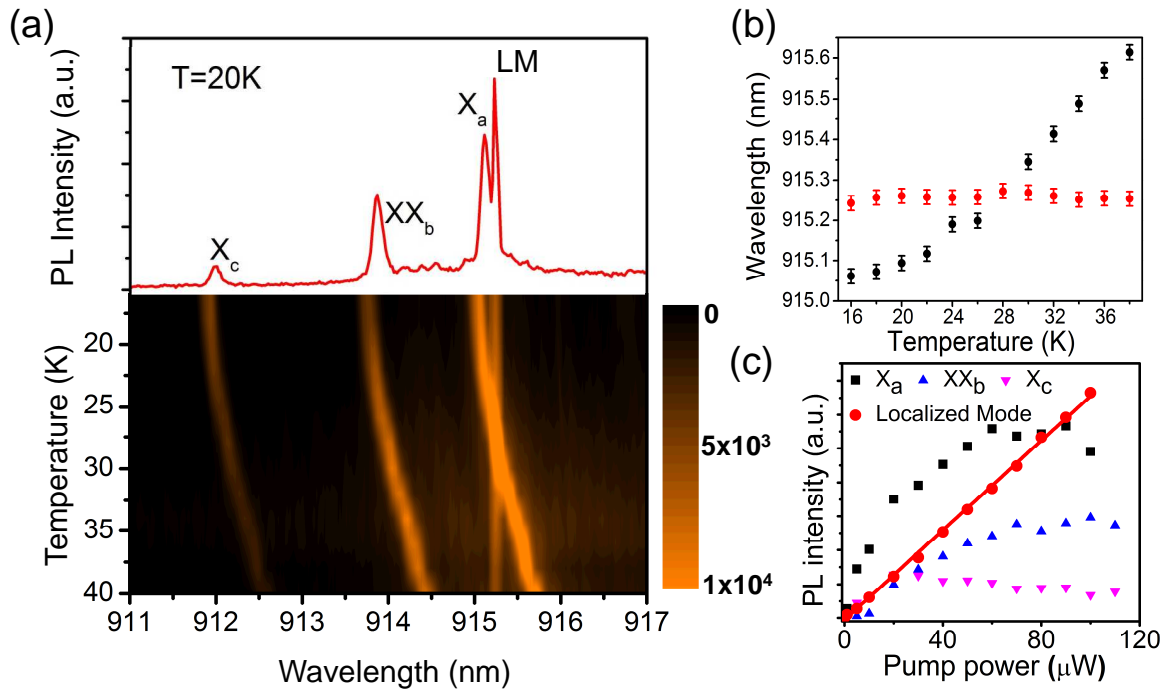

**Figure 4 | Differential temperature dependence verifications of the observed quantum dot exciton lines ($X_a$, $X_c$ and $XX_b$ lines) and localized cavity mode (LM). a,** Spectra at high excitation powers ~ 100 μW. Top panel is an example 20K spectrum. **b,** Exciton ($X_a$, black) and localized mode (LM, red) resonance peaks as a function of temperature showing crossing behavior in the weak coupling regime. **c,** the excitons and localized mode photoluminescence pump power dependence, where intensity saturation is observed for the exciton lines while a linear dependence is continually observed for the disorder-localized cavity mode.

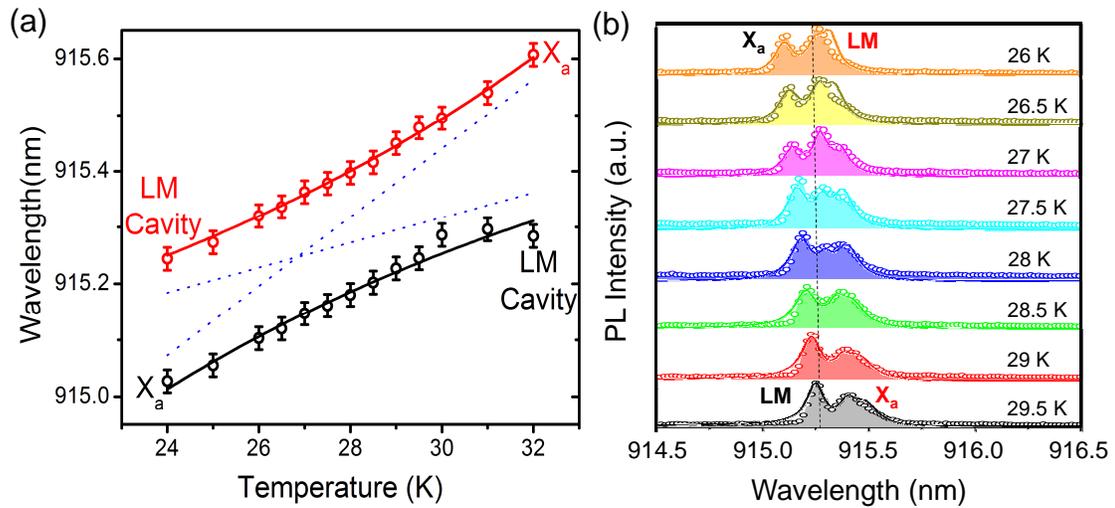

**Figure 5 | Observed coherent Rabi splitting of the disorder localized mode with single exciton state $X_a$ for different resonance-exciton detunings. a,** The two quasiparticle polariton peaks (red and black dots) from experimental spectra versus the theoretical predictions (red and black curves) as a function of measurement temperature from 24 K to 32 K. Blue dashed lines represent uncoupled cavity and exciton transitions as guidelines. **b,** Experimental spectra close to 27 K are illustrated in the open circles and the computed spectra in solid lines of the strongly coupled polaritons.

# Supplemental Information on "Strongly coupled slow-light polaritons in one-dimensional disordered localized states"

Jie Gao[1,6*], Sylvain Combrie[2], Baolai Liang[3], Peter Schmitteckert[4], Gaelle Lehoucq[2], Stephane Xavier[2], Xinan Xu[1], Kurt Busch[5], Diana L. Huffaker[3], Alfredo De Rossi[2], and Chee Wei Wong[1*]

[1]*Columbia University, New York, NY 10027 USA*

[2]*Thales Research and Technology, Palaiseau, 91767, France*

[3]*University of California at Los Angeles, Los Angeles, CA 90095, USA*

[4]*Institute of Nanotechnology, Karlsruhe Institute of Technology, Eggenstein-Leopoldshafen, 76021, Germany*

[5]*Humboldt Universität zu Berlin, Institut für Physik, AG Theoretische Optik & Photonik, and Max Born Institute, 12489 Berlin, Germany*

[6]*Department of Mechanical and Aerospace Engineering, Missouri University of Science and Technology, Rolla, MO 65409, USA*

**\*** gaojie@mst.edu; cww2104@columbia.edu

**S.I. Disorder analysis**

Uniformity and disorder in the fabricated photonic crystal lattice are analyzed with the method described in Ref. [S1]. The edge detection algorithm employed to examine the image disorder quantification involves categorizing the image into holes and the substrate region. First, we normalize the image pixel to be distributed between 0 and 1. Then, each pixel of the image is compared to an optimum threshold parameter, which is chosen based on the histogram of the pixel value of the image. Figure S1 shows one example high-resolution SEM image containing a sample photonic crystal lattice of 56 holes. A threshold value of 0.3 is chosen, and the fitted hole centers and edge shapes are shown in the white dots and curves respectively. The hole radii are found to have a mean percent error of 3.26% and the root mean square (RMS) fit error of an edge from a perfect circle is computed statistically to be approximately 3.5 nm. The RMS of the hole center deviations from a perfect lattice along the two principal directions (Figure S1) are $\sigma_x = 7.7$ nm and $\sigma_y = 2.7$ nm respectively. To study the roughness of features in the photonic crystal lattice, a fractal methodology was employed and the correlation length of 16.8 nm was computed using the parameterization of the "height-to-height" correlation function.

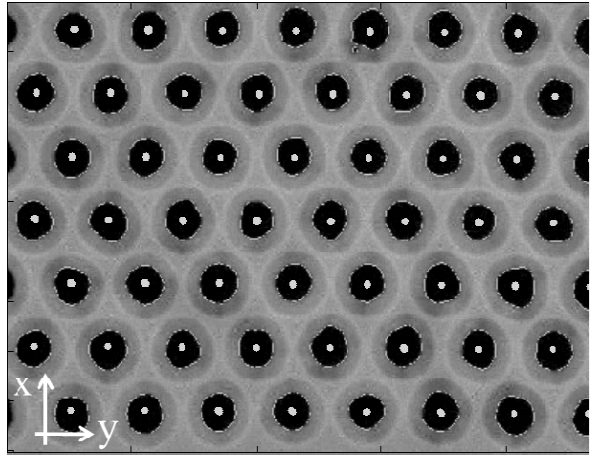

**Figure S1 | Statistical geometrical imperfections analysis of the fabricated photonic crystal lattice.** Vertices of a fitted periodic underlying lattice are shown as white dots, and edge boundaries of the fitted air holes are shown as white circles.

## S.II. Spectral function of the disordered waveguide

We consider a nearest-neighbor tight-binding Hamiltonian that describes photon propagation in an effectively one-dimensional waveguide

$$H = -J\sum_{x=1}^{M-1}(a_x^+ a_{x+1} + a_{x+1}^+ a_x) + \sum_{x=1}^{M}\varepsilon_x a_x^+ a_x \qquad (1)$$

Here $a_x^+$ and $a_x$ denote, respectively, the bosonic (photon) creation and annihilation operators at lattice site $x$ and $J$ denotes the corresponding hopping element. $M$ is the number of lattice sites, and $\varepsilon_x$ is the site energy. When measuring energies from the center of the band, the corresponding dispersion relation is $\hbar\omega_k = -2J\cos(ka)$ and we choose lattice constant $a = 1$ in our simulation without the loss the generality. To illustrate localized modes in the disordered waveguide, we examine spectral functions (imaginary part of retarded Green's function) at different wavevectors $k$ for a one-dimensional waveguide with 750 lattice sites. Disorder is introduced by taking the $\{\varepsilon_x\}$ uniformly random distributed over the interval [-$W$/2, $W$/2] [S2].

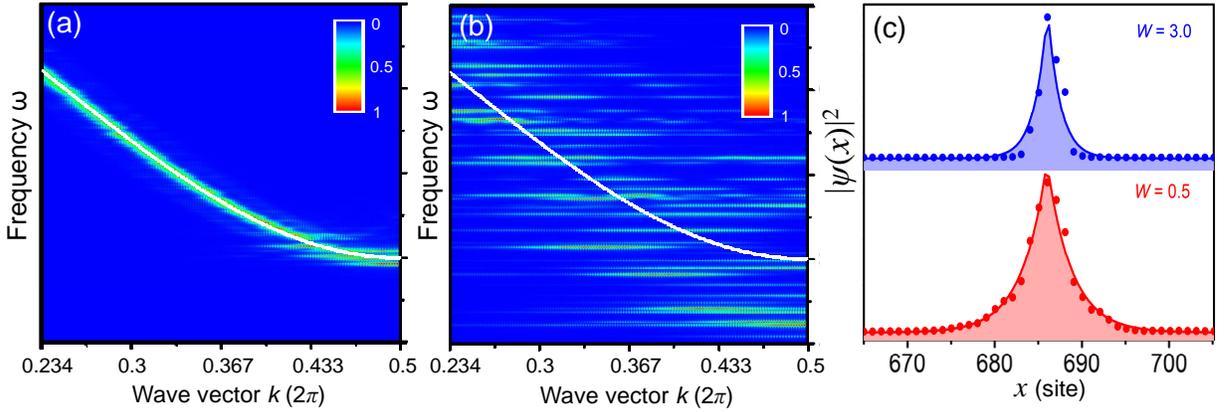

**Figure S2 | Spectral function of the disordered 1D waveguide described by the tight-binding chain model**. **a,** For a disorder strength $W$ of 0.5. **b,** For a disorder strength $W$ of 3.0. Band structure of an ideal structure is plotted in white solid line. **c,** Wavefunction (dot) and the exponential fit (curve) of an example localized mode in real space for (top) $W = 3.0$ and (bottom) $W = 0.5$.

We note that disorder strengths $W$ (also coupling strengths $V$ and transition energies $\Omega$ in section S.III) have the unit of energy as $\varepsilon_x$ and $J$, and their magnitude are normalized with respect to $J$. We also implement a Gaussian spatial filter to examine optical modes localized within a certain spatial region, which matches the case in the experiment where a photoluminescence signal from a 1 to 2 µm spot is collected by the objective lens. Figure S2(a) shows the spectral functions of a disordered waveguide with disorder strength $W = 0.5$ from mid-band to band edge. Disorder leads to a smearing of the dispersion relation which is most prominent close to the band edges. As the disorder increases to $W = 3.0$ (Figure S2(b)), the localized modes get more pronounced and even the van Hove singularity at the band edges disappears. We note that for the particular disorder configuration in Figure S2(a-b), the Gaussian spatial filter is chosen at position $x_D = 490$ with width $\sigma = 25$ where the fourth mode of the $W = 0.5$ waveguide, the sixth mode of the $W = 0.8$ waveguide, and the twentieth mode of the $W = 3.0$ waveguide can be covered by our spatial filter simultaneously. Figure S2(c) shows the shape of the wave function of one example localized mode in real space for $W = 0.5$ and $W = 3.0$. Very short localization lengths can be achieved when disorder strengths get larger as shown from the exponential fitting curves in Figure S2(c).

## S.III. Spectral function of the disordered waveguide with a quantum emitter

We extend the model to include the coupling to a quantum emitter [S3-S4]:

$$H = -J\sum_{x=1}^{M-1}(a_x^+ a_{x+1} + a_{x+1}^+ a_x) + \sum_{x=1}^{M}\varepsilon_x a_x^+ a_x + \frac{\Omega}{2}\sigma_z + V(a_{x_0}^+ \sigma_- + a_{x_0}\sigma_+) \quad (2)$$

The quantum emitter is modeled as a two-level system (described by Pauli operators) with transition frequency $\Omega/\hbar$ that is located at lattice site $x_o$ and couples with a coupling element $V$ to the modes of the photonic band. Figure S3(a-b) examine the spectral function for a quantum emitter with transition energy $\Omega = -2.05532$ at position $x_o = 488$ interacting with the fourth eigenmode for $W = 0.4$, and the spectral function for a quantum emitter with transition energy $\Omega = -2.67054$ at position $x_o = 330$ interacting with the third eigenmode for $W = 3.0$. We observe mode degeneracy at small $V$ and Rabi splitting when $V$ is at least comparable to the linewidth. The theoretically observed normal mode spectral splitting in the simulations verify the possibility of the experimentally observed strong coupling regime between a quantum emitter and localized modes in one-dimensional disordered waveguides, even in the presence of a continuum of modes. Deviating from the canonical Jaynes-Cummings ladder, this scenario can perhaps be finely described with that of modified spontaneous emission and polariton states in a Fano-like density of states.

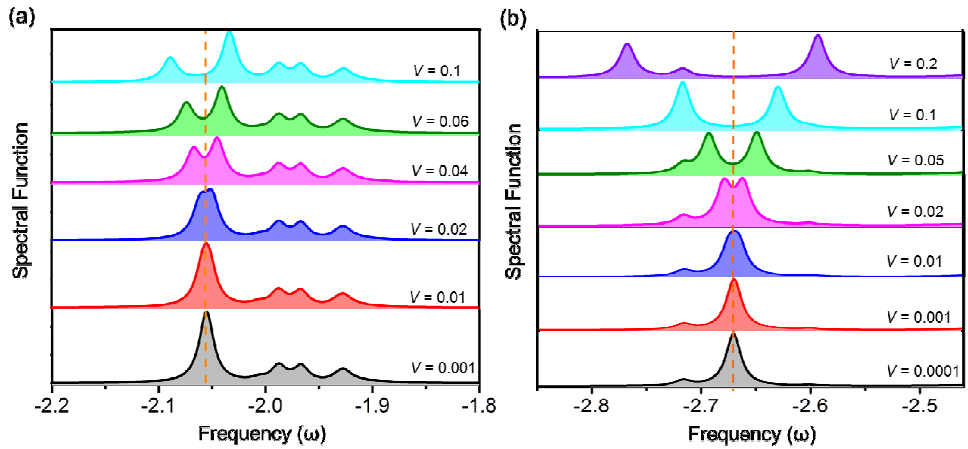

**Figure S3 | Spectral function of a disordered waveguide coupled to a quantum emitter. a,** The quantum emitter couples on resonance to one localized mode in disordered waveguide ($W = 0.4$) via coupling strengths of $V = 0.001, 0.01, 0.02, 0.04, 0.06$ and $0.1$ respectively. **b,** The quantum emitter couples on resonance to one localized mode in disordered waveguide ($W = 3.0$) via coupling strengths of $V = 0.0001, 0.001, 0.01, 0.02, 0.05, 0.1$ and $0.2$ respectively.

**Supplementary References:**

[S1] M. Skorobogatiy and G. Begin, Statistical analysis of geometrical imperfections from the images of 2D photonic crystals, *Optics Express* **13**, 2487 (2005).

[S2] P. Schmitteckert, T. Schulze, C. Schuster, P. Schwab, and U. Eckern, Anderson localization versus delocalization of interacting fermions in one dimension, *Phys. Rev. Lett.* **80**, 560 (1998).

[S3] P. Longo, P. Schmitteckert, and K. Busch, Few-photon transport in low-dimensional systems: interaction-induced radiation trapping, *Phys. Rev. Lett.* **104**, 023602 (2010).

[S4] P. Longo, P. Schmitteckert, and K. Busch, Few-photon transport in low-dimensional systems, *Phys. Rev. A* **83**, 063828 (2011).